\documentclass[12pt,twoside,english]{article}
\usepackage[T1]{fontenc}
\usepackage{graphicx}
\usepackage{amsfonts}
\usepackage{hyperref}
\usepackage[paper=a4paper,dvips,top=1.5cm,left=1.5cm,right=1.5cm,
    foot=1cm,bottom=1.5cm]{geometry}

\usepackage{todonotes}          

\usepackage{mathptmx}
\usepackage[scaled=.90]{helvet}
\usepackage{courier}
\usepackage{bookmark}

\usepackage{fancyhdr}
\usepackage[utf8]{inputenc} 
\usepackage[english]{babel}
\usepackage{array}			 
\usepackage{graphicx}	                 
\usepackage{float}			 
\usepackage{color}                       
\usepackage{mdwlist}
\usepackage{tabularx}		         
\usepackage{dcolumn}	                 
\usepackage{url}	                 
\usepackage[perpage,para,symbol]{footmisc} 
\usepackage[binary-units=true]{siunitx} 

\usepackage[all]{hypcap}	 

\definecolor{darkblue}{rgb}{0.0,0.0,0.3} 
\definecolor{darkred}{rgb}{0.4,0.0,0.0}
\definecolor{red}{rgb}{0.7,0.0,0.0}
\definecolor{lightgrey}{rgb}{0.8,0.8,0.8} 
\definecolor{grey}{rgb}{0.6,0.6,0.6}
\definecolor{darkgrey}{rgb}{0.4,0.4,0.4}
\makeatletter
\renewcommand\paragraph{\@startsection{paragraph}{4}{\z@}%
  {-3.25ex\@plus -1ex \@minus -.2ex}%
  {1.5ex \@plus .2ex}%
  {\normalfont\normalsize\bfseries}}
\makeatother

\makeatletter
\renewcommand\subparagraph{\@startsection{subparagraph}{5}{\z@}%
  {-3.25ex\@plus -1ex \@minus -.2ex}%
  {1.5ex \@plus .2ex}%
  {\normalfont\normalsize\bfseries}}
\makeatother

\usepackage[acronym, nopostdot]{glossaries}
\glsdisablehyper
\makeglossaries

\newacronym{PLS}{PLS}{Physical Layer Security}
\newacronym{MIMO}{MIMO}{Multiple-Input-Multiple-Output}
\newacronym{MRT}{MRT}{Maximum Ratio Transmission}
\newacronym{ZF}{ZF}{Zero-Forcing}
\newacronym{AN}{AN}{Artificial Noise}
\newacronym{CSI}{CSI}{Channel State Information}
\newacronym{SINR}{SINR}{Signal-to-Interference-plus-Noise Ratio}
\newacronym{SOP}{SOP}{Secrecy Outage Probability}
\newacronym{EE}{EE}{Energy Efficiency}
\newacronym{BS}{BS}{Base Station}
\newacronym{SNR}{SNR}{Signal-to-Noise Ratio}
\newacronym{IoT}{IoT}{Internet of Things}
\newacronym{6G}{6G}{Sixth Generation}
\newacronym{5G}{5G}{Fifth Generation}
\newacronym{CDF}{CDF}{Cumulative Distribution Function}
\newacronym{MMSE}{MMSE}{Minimum Mean Square Error}
\newacronym{SVD}{SVD}{Singular Value Decomposition}
\newacronym{AWGN}{AWGN}{Additive White Gaussian Noise}
\newacronym{RIS}{RIS}{Reconfigurable Intelligent Surface}
\newacronym{RF}{RF}{Radio Frequency}
\newacronym{TDD}{TDD}{Time Division Duplex}
\newacronym{OFDM}{OFDM}{Orthogonal Frequency Division Multiplexing}
\newacronym{QoS}{QoS}{Quality of Service}

\title{Physical Layer Security in Massive MIMO: Challenges and Open Research Directions Against Passive Eavesdroppers}
\author{
        \textsc{Nipun Agarwal}
            \qquad
        \normalsize
            \texttt{nipuna}
            \texttt{@kth.se}
}


\begin{document}

\maketitle

\begin{abstract}
\label{sec:abstract}

Massive \gls{MIMO} has become a crucial enabling technology for 5G and beyond, providing previously unheard-of increases in energy and spectrum efficiency. It is still difficult to guarantee secure communication in these systems, particularly when it comes to passive eavesdroppers whose base station is unaware of their channel state information. By taking advantage of the inherent randomness of wireless channels, \gls{PLS} offers a promising paradigm; however, its efficacy in massive \gls{MIMO} is heavily reliant on resource allocation and transmission strategies. In this work, the performance of secure transmission schemes, such as \gls{MRT}, \gls{ZF} and \gls{AN}-aided beamforming, is examined when passive eavesdroppers are present. This work will use extensive Monte Carlo simulations to assess important performance metrics such as energy efficiency, secrecy outage probability, and secrecy sum rate under different system parameters (e.g., number of antennas, \gls{SNR}, power allocation). The results aim to provide comparative insight into the strengths and limitations of different \gls{PLS} strategies and to highlight open research directions to design scalable, energy-efficient, and robust secure transmission techniques in future 6G networks.
\end{abstract}

\vspace{0.5em}
\noindent\textit{\textbf{Keywords -}} Physical Layer Security, Massive MIMO, Zero-Forcing, Artificial Noise, Secrecy Rate, Channel Estimation.

\selectlanguage{english}
\tableofcontents

\clearpage
\section{Introduction}
\label{sec:introduction}

The exponential growth of wireless communications and the emergence of \gls{IoT} applications have created unprecedented demands for both high-performance and secure transmission systems. With over 50 billion connected devices anticipated by 2030, traditional cryptographic approaches face significant challenges from quantum computing threats and computational complexity constraints. \gls{PLS} emerges as a fundamental paradigm shift that exploits the inherent randomness and unique characteristics of wireless channels to achieve information-theoretic secrecy without relying solely on computational assumptions \cite{wyner1975wire,cover2006elements}.

Massive \gls{MIMO} technology, characterised by deploying hundreds of antennas at base stations to serve multiple users simultaneously, has become the cornerstone of \gls{5G} networks and is anticipated to play an even more crucial role in \gls{6G} systems \cite{marzetta2010noncooperative,larsson2014massive}. Following established system configuration parameters by Ngo et al. \cite{ngo2020massive} and Björnson et al. \cite{bjornson2021energy}, this research employs $K=4$ users served by base stations equipped with $M \in \{32,64,128,256\}$ antennas. The large antenna arrays enable unprecedented spatial resolution, enhanced beamforming capabilities, and improved energy efficiency through favourable propagation conditions \cite{rusek2013scaling}.

\subsection{Literature Study}
\label{subsec:related_work_extended}

The theoretical foundations of \gls{PLS} originate from Wyner's wiretap channel and were developed further in standard textbooks on information theory; these works formalise secrecy capacity as the difference between the legitimate receiver's mutual information and that of the eavesdropper \cite{wyner1975wire,cover2006elements}. This information-theoretic perspective motivated early practical techniques that exploit channel randomness to provide secrecy guarantees independent of computational assumptions. In parallel, the growth of wireless infrastructure and multi-antenna technologies prompted a large body of work examining how spatial processing can be used to shape secrecy performance in practical fading environments \cite{oggier2011secrecy,goel2008guaranteeing}.

\gls{AN} methods were introduced to degrade the eavesdropper's reception while minimally affecting intended users by placing noise in the transmitter's nullspace; these ideas have been extended to multi-user \gls{MIMO}, cooperative jamming, and joint beamforming/AN optimisation frameworks \cite{goel2008guaranteeing,chen2019artificial,si2020cooperative}. The literature shows that \gls{AN} can be especially effective when the transmitter has accurate \gls{CSI} about legitimate channels and when the eavesdropper does not lie in the same spatial directions as legitimate users; conversely, \gls{AN}'s benefits diminish and even reverse when \gls{CSI} is poor or when power resources are limited \cite{wang2023robust,zheng2023artificial}.

Massive \gls{MIMO} changes many aspects of the secrecy problem. As the number of base-station antennas grows, phenomena such as channel hardening and favourable propagation simplify certain design tasks and increase the degrees of freedom available for spatial nulling and beamforming \cite{marzetta2010noncooperative,rusek2013scaling,ngo2020massive}. However, practical massive \gls{MIMO} systems must confront limitations that directly impact secrecy: pilot contamination in \gls{TDD} systems, finite coherence times that limit pilot length, and non-ideal hardware that increases per-antenna power consumption. Several surveys and focused studies examine the interplay between antenna scaling and security, arguing that massive arrays can improve secrecy but only when \gls{CSI} acquisition and pilot assignment are carefully managed \cite{bjornson2017massive,bjornson2021energy,oggier2011secrecy}.


Frequency-dependent propagation has also been shown to influence \gls{PLS}. The mmWave bands offer highly directional beams and sparser multipath, which can enhance spatial discrimination between legitimate receivers and eavesdroppers; however, mmWave also presents higher path loss, beam alignment sensitivity and hardware cost implications that must be included in any realistic comparison \cite{maccartney2022outdoor,akdeniz2020millimeter,rangan2021millimeter}. Empirical and simulation studies indicate that mmWave can produce better secrecy under favourable geometry, but that these advantages are not universal and depend on environment, beamforming accuracy and mobility \cite{maccartney2022outdoor,bjornson2021energy}.

Energy efficiency and end-to-end cost considerations have become increasingly important in the design of secure systems. Recent massive \gls{MIMO} studies incorporate circuit-power-per-antenna models and show that gains in spectral efficiency or secrecy can be offset by rising circuit power as antenna count increases \cite{abou2021energy,bjornson2021energy}. This energy–security interaction means that a scheme that appears optimal when only transmit power is considered (for example, heavy use of AN) may be suboptimal once realistic hardware energy costs are included; the study therefore includes per-antenna circuit power in the energy-efficiency metric to produce more deployment-relevant insights \cite{abou2021energy,li2024quantum}.


Despite the large and growing literature, comparative analyses that simultaneously incorporate \gls{MMSE}-based \gls{CSI} uncertainty, per-antenna circuit power, multi-antenna passive eavesdroppers, and frequency-dependent channel models remain rare. Prior studies often examine a subset of these aspects in isolation e.g., \gls{AN} effectiveness with perfect \gls{CSI}, or antenna scaling without energy models making it difficult to form robust deployment recommendations. Moreover, while some recent papers evaluate massive \gls{MIMO} \gls{PLS} empirically, many do not provide full statistical reporting (confidence intervals and effect sizes), which hinders reproducibility and industry adoption \cite{zeng2021monte,bjornson2017massive}.

This report addresses these gaps by offering a unified and statistically rigorous comparison of \gls{MRT}, \gls{ZF}, \gls{AN}-aided \gls{MRT}, and a robust regularised precoder across a wide parameter space (antenna counts, \gls{SNR}, \gls{AN} power split, \gls{CSI} error levels, and eavesdropper capability). By combining dual-band propagation modelling (sub-6\, GHz and mmWave) with \gls{MMSE}-based \gls{CSI} uncertainty and per-antenna circuit power, the framework yields operationally relevant recommendations and identifies regimes where conventional intuitions (e.g., ``\gls{AN} always helps'') fail. The results, therefore, offer both theoretical insight and practical guidance for system designers and operators looking to deploy secure massive \gls{MIMO} in near-term and future networks \cite{ngo2020massive,bjornson2021energy,goel2008guaranteeing}.

\subsection{Theoretical Framework}
\label{subsec:framework}

The theoretical foundation of this work builds upon Wyner's seminal wiretap channel work \cite{wyner1975wire}, which established the fundamental principles of information-theoretic security. The secrecy capacity framework forms the mathematical basis for the analysis:
\begin{equation}
R_s = \max_{p(x)} [I(X; Y) - I(X; Z)] = C_{main} - C_{wiretap}
\label{eq:wyner_capacity}
\end{equation}
where the maximum secrecy rate $R_s$ is the largest difference between the mutual information $I(X;Y)$ (between transmitter and intended receiver) and $I(X;Z)$ (between transmitter and eavesdropper). Equivalently, it is the main channel capacity $C_{\rm main}$ minus the wiretap (eavesdropper) channel capacity $C_{\rm wiretap}$. Here $p(x)$ is the transmit distribution. The interpretation is that one can reliably transmit at rate $C_{\rm main}$.
\\
\newline
\textbf{Complex Gaussian Channel Coefficient Generation:} Following the mathematical framework, channel coefficients are generated using statistical methods. This defines a complex Gaussian channel coefficient $h_{m,k}$ (from BS antenna $m$ to user $k$) in terms of real ($h^{(r)}$) and imaginary ($h^{(i)}$) parts. The prefactor $1/\sqrt{2}$ ensures unit variance: if $h^{(r)}$ and $h^{(i)}$ are each real $\mathcal{N}(0,1)$, then $h_{m,k}\sim\mathcal{CN}(0,1)$. The small-scale fading is modelled as Rayleigh, and this formula shows how to generate it from two independent normal random variables. The assumption is that the real and imaginary parts are independent with equal variance:
\begin{equation}
h_{m,k} = \frac{1}{\sqrt{2}}\left(h_{m,k}^{(r)} + jh_{m,k}^{(i)}\right)
\label{eq:channel_gaussian}
\end{equation}
(see Appendix~\ref{der:1} for a complete derivation).
\\
\newline
\textbf{Path Loss Channel Model:} The complete channel model incorporating large-scale fading effects follows:
\begin{equation}
\mathbf{H}_k = \sqrt{\beta_k} \tilde{\mathbf{H}}_k
\label{eq:channel_pathloss}
\end{equation}
where $\tilde{\mathbf{H}}_k$ is an $M \times 1$ vector whose entries are i.i.d. $\mathcal{CN}(0,1)$ (as per Eq. \ref{eq:channel_gaussian} for each antenna), representing the small-scale fading from the BS to user $k$. The scalar $\beta_k$ is the large-scale fading (path loss and shadowing) for user $k$. Physically, $\beta_k<1$ accounts for attenuation (e.g. $\beta_k = d_k^{-\alpha}$ for pathloss exponent $\alpha$). This model assumes a block-fading channel with independent Rayleigh fading scaled by $\sqrt{\beta_k}$. All entries of $\tilde{\mathbf{H}}_k$ are uncorrelated (no spatial correlation).
(see Appendix~\ref{der:2} for a complete derivation).

\subsection{Research Questions and Hypotheses}
\label{subsec:research_questions}

This research addresses fundamental questions that have remained inadequately explored in the literature:
\\
\newline
\textbf{Primary Research Question:} How do different secure transmission schemes (\gls{MRT}, \gls{ZF}, MRT+\gls{AN}, Robust) perform in massive \gls{MIMO} systems against passive eavesdroppers under realistic conditions with imperfect \gls{CSI}, hardware constraints, and varying threat models?
\\
\newline
\textbf{Secondary Research Questions:}
\begin{enumerate}
\item Which precoding scheme provides the optimal balance between secrecy rate, outage probability, and energy efficiency across comprehensive parameter variations?
\item How does base station antenna scaling impact security performance, and what are the practical deployment thresholds?
\item What are the quantitative security advantages of sub-6 GHz versus mmWave frequency bands under realistic propagation conditions?
\item How should power be optimally allocated between information transmission and artificial noise injection across different operating scenarios?
\item What is the statistical reliability and practical significance of performance differences between schemes?
\item How do eavesdropper capabilities (antenna count, processing sophistication) impact system security across different schemes?
\item What are the computational complexity trade-offs and their practical implementation implications?
\end{enumerate}

\textbf{Research Hypotheses with Quantitative Predictions:}
\begin{enumerate}
\item[\textbf{H1:}] \gls{ZF} precoding will achieve $>10\%$ superior secrecy performance compared to other schemes due to interference suppression capabilities, but with higher computational complexity.
\item[\textbf{H2:}] Antenna scaling will provide logarithmic improvements in secrecy performance up to a certain ratio, beyond which diminishing returns occur due to pilot contamination effects.
\item[\textbf{H3:}] mmWave frequency bands will demonstrate superior physical layer security due to enhanced spatial isolation, despite higher noise figures.
\item[\textbf{H4:}] Optimal power allocation will follow non-monotonic relationships with peak performance at $\rho = 0.6\text{-}0.7$ for \gls{AN} schemes, varying with channel conditions and threat severity.
\item[\textbf{H5:}] Performance differences will be statistically significant with effect sizes $>0.8$ and confidence levels $>95\%$ across most operating conditions.
\end{enumerate}

\section{Method}
\label{sec:method}

This research employs an exhaustive quantitative methodology combining rigorous theoretical analysis, complete mathematical derivations, Monte Carlo simulations, and extensive statistical validation. The methodology integrates realistic channel models, practical hardware constraints, and statistical frameworks to ensure reliable and generalizable results \cite{stirling2025reproducibility}.

\subsection{System Model and Mathematical Framework}
\label{subsec:system_model}

\textbf{System Architecture and Configuration:} Following established parameters by Ngo et al. \cite{ngo2020massive}, the study considers a downlink massive \gls{MIMO} system where a base station with $M$ base station antennas serves $K=4$ single-antenna legitimate users in the presence of a passive eavesdropper with $N_e \in \{1,2,4\}$ antennas. The system operates across dual frequency bands: sub-6 GHz at 3.5 GHz and mmWave at 28 GHz, following 3GPP TR 38.901 \cite{3gpp2020study} and Rangan et al. \cite{rangan2021millimeter}.
\\
\newline
\textbf{Channel State Information Modelling :} Practical systems suffer from imperfect \gls{CSI} due to pilot-based channel estimation limitations. The study models \gls{CSI} imperfections through the \gls{MMSE} estimation framework with error variances $\epsilon_{CSI} \in \{0.01, 0.10, 0.30\}$ representing near-perfect to severely degraded estimation quality. The imperfect channel estimate is:
\begin{equation}
\hat{\mathbf{H}}_k = \mathbf{H}_k + \mathbf{E}_k,
\label{eq:imperfect_csi}
\end{equation}
This expresses that the base station’s estimated channel $\hat{\mathbf{H}}_k$ differs from the true channel $\mathbf{H}_k$ by an error $\mathbf{E}_k$ (for user $k$). The model assumes estimation via uplink pilots, so $\mathbf{E}_k$ captures noise and any pilot contamination. Under \gls{MMSE} estimation:
\begin{align}
\mathbf{Y}_p &= \sqrt{p_p}\mathbf{H}_k \boldsymbol{\phi}_k + \mathbf{N}_p, \\
\hat{\mathbf{H}}_k &= \frac{\sqrt{p_p}}{p_p|\boldsymbol{\phi}_k|^2 + \sigma_n^2}\mathbf{Y}_p\boldsymbol{\phi}_k^H,
\label{eq:mmse_estimate}
\end{align}
This models the received pilot signal $\mathbf{Y}_p$ at the base station for user $k$. Here $p_p$ is the pilot transmit power, $\boldsymbol{\phi}_k$ is the pilot sequence (size matching coherence length), and $\mathbf{N}_p$ is noise. Essentially, $\mathbf{H}_k,\boldsymbol{\phi}_k$ is how the pilot is observed through the channel. This linear model assumes \gls{TDD} and orthogonal pilots. This is the \gls{MMSE} channel estimate at the base station for user $k$. Here $|\boldsymbol{\phi}_k|^2$ is the norm of the pilot sequence, and $\sigma_n^2$ is the noise variance. If $|\boldsymbol{\phi}_k|^2=1$ (unit pilots) and ignoring $\sigma_n^2$, this reduces to $\hat{\mathbf{H}}_k\approx \mathbf{H}_k$. The exact expression comes from standard \gls{MMSE} estimation under Gaussian noise.

The expanded \gls{MMSE} estimate is:
\begin{equation}
\label{eq: equation_7}
\hat{\mathbf{H}}_k = \frac{p_p}{p_p + \sigma_n^2}\mathbf{H}_k + 
\frac{\sqrt{p_p}\sigma_n^2}{p_p + \sigma_n^2}\mathbf{N}_p\boldsymbol{\phi}_k^H,
\end{equation}
It shows that the estimate is a weighted sum of the true channel and a noise term. The coefficient $\frac{p_p}{p_p+\sigma_n^2}$ multiplies the true channel, reducing its magnitude by estimation dilution, while the second term is residual noise (scaled by $\frac{\sqrt{p_p},\sigma_n^2}{p_p+\sigma_n^2}$). The decomposition makes clear that $\hat{\mathbf{H}}_k$ is unbiased but with an error variance proportional to $\sigma_n^2$.

\begin{equation}
\mathbb{E}[\mathbf{E}_k\mathbf{E}_k^H] = \frac{\sigma_n^2}{p_p + \sigma_n^2}\mathbf{I}_M.
\end{equation}
This equation gives the error covariance: it is diagonal $\propto\mathbf{I}_M$ with variance $\sigma_n^2/(p_p+\sigma_n^2)$ per antenna. Each antenna’s channel estimate error has variance $\sigma_n^2/(p_p+\sigma_n^2)$, assuming normalised pilot power. This follows from Eq. \ref{eq: equation_7} since $\mathbf{N}_p\boldsymbol{\phi}_k^H$ is noise and uncorrelated across antennas. The assumption is i.i.d. noise, leading to this simple form.

This framework ensures realistic modelling of implementation constraints that significantly impact security performance.

\subsection{Beamforming Algorithm Implementation:} 
\label{subsec:beamforming_algo}
This work implements and analyses four distinct secure transmission schemes:
\\
\newline
\textbf{1. Maximum Ratio Transmission (MRT):} The \gls{MRT} precoding vector maximizes received \gls{SNR} at the intended user:
\begin{equation}
\label{eq:mrt_final}
\mathbf{w}_k^{MRT} = \frac{\hat{\mathbf{h}}_k}{\|\hat{\mathbf{h}}_k\|}.
\end{equation}
(see Appendix~\ref{der:3} for a complete derivation), here $\hat{\mathbf{h}}_k$ is the (estimated) channel vector from the base station to user $k$ (size $M\times1$). This choice of $\mathbf{w}_k$ (normalised to unit norm) aims to maximise the received signal power at user $k$. No weight is given to suppressing interference; each beam is simply pointed along the channel direction. The normalization $|\hat{\mathbf{h}}_k|$ ensures $|\mathbf{w}_k|=1$. One assumes $\mathbf{w}_k$ is applied with power $\sqrt{P_s}$. The assumption here is a single-antenna user, so \gls{MRT} is optimal for \gls{SNR} on its link.
\\
\newline
\textbf{2. Zero-Forcing (ZF) Precoding:} \gls{ZF} precoding eliminates inter-user interference via orthogonal precoding design:
\begin{align}
\mathbf{W}^{ZF} &= \hat{\mathbf{H}}(\hat{\mathbf{H}}^H\hat{\mathbf{H}})^{-1}, \\
\hat{\mathbf{H}}^H\mathbf{W}^{ZF} &= \mathbf{I}_K,
\end{align}
here, $\hat{\mathbf{H}} = [\hat{\mathbf{h}}_1, \dots, \hat{\mathbf{h}}_K]$ is the $M\times K$ estimated channel matrix to all $K$ users. The product $(\hat{\mathbf{H}}^H\hat{\mathbf{H}})^{-1}$ is a $K\times K$ inverse assuming $M\ge K$ and full rank. Thus $\mathbf{W}^{\rm ZF}$ is $M\times K$. Each column $\tilde{\mathbf{w}}_k$ of $\mathbf{W}^{\rm ZF}$ is an unnormalized beamformer for user $k$. The idea is that by choosing $\mathbf{W}^{\rm ZF}$ this way, it ensures $\hat{\mathbf{H}}^H \mathbf{W}^{\rm ZF} = \mathbf{I}_K$, meaning each user sees no inter-user interference. It states that after precoding, the effective channel matrix from base station to users is (approximately) the identity, so each user’s signal arrives cleanly without contributions from other users’ beams.

The normalised precoding vectors are:
\begin{equation}
\mathbf{w}_k^{ZF} = \frac{\tilde{\mathbf{w}}_k}{\|\tilde{\mathbf{w}}_k\|},
\end{equation}

where $\tilde{\mathbf{w}}_k$ is the $k$-th column of $(\hat{\mathbf{H}}^H\hat{\mathbf{H}})^{-1}$. This condition requires $M \ge K$ and full column rank of $\hat{\mathbf{H}}$.
\\
\newline
\textbf{3. Robust Precoding with Uncertainty Management:} To enhance robustness against \gls{CSI} uncertainties, regularised precoding incorporates estimation error statistics:
\begin{equation}
\mathbf{w}_k^{Robust} = \frac{\hat{\mathbf{h}}_k}{\sqrt{\|\hat{\mathbf{h}}_k\|^2 + \alpha_k}},
\label{eq:robust_precoding}
\end{equation}

It is similar to \gls{MRT} (pointing along $\hat{\mathbf{h}}_k$) but with a regularisation term $\alpha_k$. Here $\alpha_k$ is chosen based on the \gls{CSI} error variance for user $k$. The idea is that when \gls{CSI} is imperfect, this avoids over-emphasising the channel estimate, thus limiting the impact of estimation error. The denominator can be seen as reducing the gain on the channel; if $\alpha_k=0$, it reduces to \gls{MRT}. This form implicitly assumes scalar (per-user) regularisation. It arises from derivations (see Appendix~\ref{der:5} of worst-case \gls{SINR} maximisation under Gaussian error. This mitigates performance degradation due to imperfect \gls{CSI}.
\\
\newline
\textbf{4. Artificial Noise (AN) Design with Null-Space Projection:} Artificial noise enhances security by degrading eavesdropper channels while minimising impact on legitimate users \cite{goel2008guaranteeing}. Let $\hat{\mathbf{H}} = \mathbf{U}\boldsymbol{\Sigma}\mathbf{V}^H$ be the \gls{SVD} of the estimated channel, and partition $\mathbf{U}=[\mathbf{U}_1,\mathbf{U}_0]$ with $\mathbf{U}_1 \in \mathbb{C}^{M\times K}$ spanning the column space and $\mathbf{U}_0 \in \mathbb{C}^{M\times (M-K)}$ spanning the null space. The artificial noise covariance is:
\begin{equation}
\mathbf{Q}_{AN} = \mathbf{U}_0 \mathbf{U}_0^H,
\end{equation}
which places artificial noise in the null space of legitimate user channels, ensuring no interference to intended signals. The matrix $\mathbf{U}_0$ satisfies:
\begin{align}
\hat{\mathbf{H}}\mathbf{U}_0 &= \mathbf{0} \quad \text{(orthogonality condition)},\\
\mathbf{U}_0^H\mathbf{U}_0 &= \mathbf{I}_{M-K} \quad \text{(normalization condition)}.
\end{align}
This guarantees that artificial noise degrades the eavesdropper channels while preserving the signal to intended users. The underlying assumption is $M>K$, so a non-trivial nullspace exists.

\subsection{Simulation Methodology and Statistical Framework}
\label{subsec:simulation}

\textbf{Experimental Design:} The simulation methodology employs a full factorial design across extensive parameter variations to ensure comprehensive coverage and statistical reliability. The experimental framework follows established Monte Carlo approaches with 1000 realisations, providing statistical power $>0.8$ for detecting meaningful performance differences \cite{zeng2021monte}. All random processes use a fixed seed (42) for reproducibility \cite{stirling2025reproducibility}, using Python v3.13.3 as the simulation platform.

\begin{table}[H]
\centering
\caption{System Configuration Parameters}
\label{tab:system_config}
\begin{tabular}{|p{4cm}|p{3cm}|p{6cm}|}
\hline
\textbf{Parameter} & \textbf{Value} & \textbf{Reference Source} \\
\hline
Number of users & $K = 4$ & Ngo et al. \cite{ngo2020massive} \\
\hline 
Base station antennas \& $M \in \{32,64,128,256\}$ & Björnson et al. \cite{bjornson2021energy} \\
\hline
Eavesdropper antennas & $N_e \in \{1,2,4\}$ & He et al. \cite{he2022physical} \\
\hline
\gls{CSI} error variance & $\{0.01,0.10,0.30\}$ & Wang et al. \cite{wang2023robust} \\
\hline
Monte Carlo runs & 1000 & Zeng et al. \cite{zeng2021monte} \\
\hline
Random seed & 42 & Stirling et al. \cite{stirling2025reproducibility} \\
\hline
\end{tabular}
\end{table}

\begin{table}[H]
\centering
\caption{Radio-Frequency and Power Parameters}
\label{tab:rf_params}
\begin{tabular}{|p{4cm}|p{5cm}|p{6cm}|}
\hline
\textbf{Parameter} & \textbf{Value} & \textbf{Reference Source} \\
\hline
Sub-6 GHz frequency & 3.5 GHz & 3GPP TR 38.901 \cite{3gpp2020study} \\
\hline
mmWave frequency & 28 GHz & Rangan et al. \cite{rangan2021millimeter} \\
\hline
Path loss (users) & 1.0 (sub-6GHz) / 0.3 (mmWave) & MacCartney et al. \cite{maccartney2022outdoor} \\
\hline
Path loss (eavesdroppers) & 0.5 (sub-6GHz) / 0.8 (mmWave) & MacCartney et al. \cite{maccartney2022outdoor} \\
\hline
Noise figure & 7 dB (sub-6GHz), 9 dB (mmWave) & Akdeniz et al. \cite{akdeniz2020millimeter} \\
\hline
\gls{SNR} range & $\{-10,0,10,20,30,40,50\}$ dB & Shi et al. \cite{shi2024secrecy} \\
\hline
Power split factor & $\rho \in \{0,0.2,0.4,0.6,0.8,1.0\}$ & Zheng et al. \cite{zheng2023artificial} \\
\hline
Target secrecy rate & 0.5 bits/s/Hz & Liu et al. \cite{liu2022secrecy} \\
\hline
Circuit power per antenna & 0.1 W $\times$ M & Abou-Rjeily et al. \cite{abou2021energy} \\
\hline
\end{tabular}
\end{table}

The parameter selection ensures comprehensive coverage of practical deployment scenarios while maintaining computational tractability. The choice of $K=4$ users represents typical small-cell scenarios, while the antenna range $M \in \{32,64,128,256\}$ covers current massive \gls{MIMO} deployments through future \gls{6G} systems.

\subsection{Performance Metrics and SINR Formulation}
\label{subsec:metrics}

\textbf{Transmitted signal:}
\begin{equation}
\mathbf{x} = \sum_{k=1}^K \sqrt{P_s}\,\mathbf{w}_k\,s_k + \mathbf{n}_{AN},
\label{eq:transmitted_signal2}
\end{equation}

This is the transmitted baseband signal vector from the $M$-antenna base station. It consists of $K$ users’ symbols $s_k$ (assumed unit energy) each beamformed by $\mathbf{w}_k$, scaled by $\sqrt{P_s}$ where $P_s=\rho P{\rm total}$ is the per-user signal power. The term $\mathbf{n}_{AN}$ is the artificial noise vector with covariance $\mathbf{Q}_{AN}$ and power $P_{AN}=(1-\rho)P_{\rm total}$ (so total transmit power is $P_{\rm total}$). This model assumes linear superposition of data and \gls{AN}.
\\
\newline
\textbf{Received signal at legitimate user $k$:}
\begin{equation}
y_k = \mathbf{h}_k^H \mathbf{w}_k \, s_k + \sum_{j \ne k} \mathbf{h}_k^H \mathbf{w}_j\, s_j + \mathbf{h}_k^H \mathbf{n}_{AN} + n_k,
\label{eq:received_signal2}
\end{equation}

This is the received signal at legitimate user $k$. The first term is the desired signal, the second term is inter-user interference from other beams, the third term is the effect of artificial noise on user $k$ (which ideally is zero if $\mathbf{w}_k$ and $\mathbf{n}_{AN}$ are orthogonal) and the last is receiver noise $n_k\sim\mathcal{CN}(0,\sigma_n^2)$. Under ideal nullspace AN, $\mathbf{h}_k^H \mathbf{n}_{AN}=0$, but in practice, with estimation error, it may not vanish completely. The formula assumes linear beamforming and additive white Gaussian noise.
\\
\newline
\textbf{SINR at legitimate user $k$:}
\begin{equation}
\gamma_k = \frac{P_s\,|\mathbf{h}_k^H\mathbf{w}_k|^2}
{\sum_{j\ne k} P_s\,|\mathbf{h}_k^H\mathbf{w}_j|^2 + \mathbf{h}_k^H\mathbf{Q}_{AN}\mathbf{h}_k + \sigma_n^2}.
\label{eq:user_sinr2}
\end{equation}

This is the signal-to-interference-plus-noise ratio (SINR) at user $k$. The numerator is the signal power, while the denominator sums: (i) interference from other users’ beams $\sum_{j\ne k} P_s\,|\mathbf{h}_k^H\mathbf{w}_j|^2$, (ii) the power of artificial noise seen by user $k$ given by $\mathbf{h}_k^H \mathbf{Q}_{AN}\mathbf{h}_k$ (which is ideally zero for perfect nullspace AN and channel knowledge), and (iii) thermal noise $\sigma_n^2$. In deriving this, one assumes independent noise and normalised symbols. If \gls{ZF} precoding is used, the interference term can be zero under perfect \gls{CSI}, simplifying the \gls{SINR}.
\\
\newline
\textbf{SINR at eavesdropper antenna $e$ (attempting to decode user $k$):}
\begin{equation}
\label{eq:eve_sinr2}
\gamma_{k,e} = \frac{P_s\,|\mathbf{g}_{k,e}^H\mathbf{w}_k|^2}
{\sum_{j\ne k} P_s\,|\mathbf{g}_{k,e}^H\mathbf{w}_j|^2 + P_{AN}\mathbf{g}_{k,e}^H\mathbf{Q}_{AN}\mathbf{g}_{k,e} + \sigma_n^2}.
\end{equation}

where, $\mathbf{g}_{k,e}$ is the channel from the base station to eavesdropper antenna $e$ for user $k$’s signal. The formula mirrors Eq. \ref{eq:user_sinr2}: numerator is desired power via $w_k$, denominator includes interference from other beams (the eaves hears all users), plus the artificial noise term $P_{AN} \mathbf{g}_{k,e}^H \mathbf{Q}_{AN}\mathbf{g}_{k,e}$. Here, the study assumes the eavesdropper can combine all its $N_e$ antennas (taking the maximum $\gamma{k,e}$ over $e$ in Eq. \ref{eq:secrecy_rate2}. The model assumes the eavesdropper does not know the signals or codes, so it treats interference and \gls{AN} as noise. This expression assumes the eavesdropper is passive with an unknown channel at the base station, so $\mathbf{g}_{k,e}$ is independent and possibly weaker (different $\beta$) than $\mathbf{h}_k$.
\\
\newline
\textbf{Per-user secrecy rate:}
\begin{equation}
R_s^{(k)} = \max\{0,\; \log_2(1+\gamma_k) - \max_e \log_2(1+\gamma_{k,e})\}.
\label{eq:secrecy_rate2}
\end{equation}

This defines the achieved secrecy rate for user $k$. It is the positive part of the difference between the user’s achievable rate $\log_2(1+\gamma_k)$ and the maximum rate any eavesdropper antenna (indexed by $e$) could decode ($\max_e \log_2(1+\gamma_{k,e})$). If the eavesdropper channel is better for some antenna $e$, it effectively limits secure transmission. The $\max\{0,\cdot\}$ ensures a non-negative rate (if the eavesdropper is too strong, the secrecy rate is zero). This follows Wyner’s construction for multi-antenna wiretap channels. It assumes Gaussian signalling and worst-case eavesdropper combining.

\section{Results and Analysis}
\label{sec:results}

This section presents results from extensive Monte Carlo simulations encompassing over 1000 runs across all parameter combinations. The analysis integrates theoretical predictions with empirical validation through detailed performance plots, providing exhaustive insights into secure transmission scheme performance under realistic massive \gls{MIMO} conditions.

\subsection{Performance Characterisation and Statistical Analysis}
\label{subsec:statistical_analysis}

\begin{figure}
\centering
\includegraphics[width=0.8\textwidth,height=0.5\linewidth]{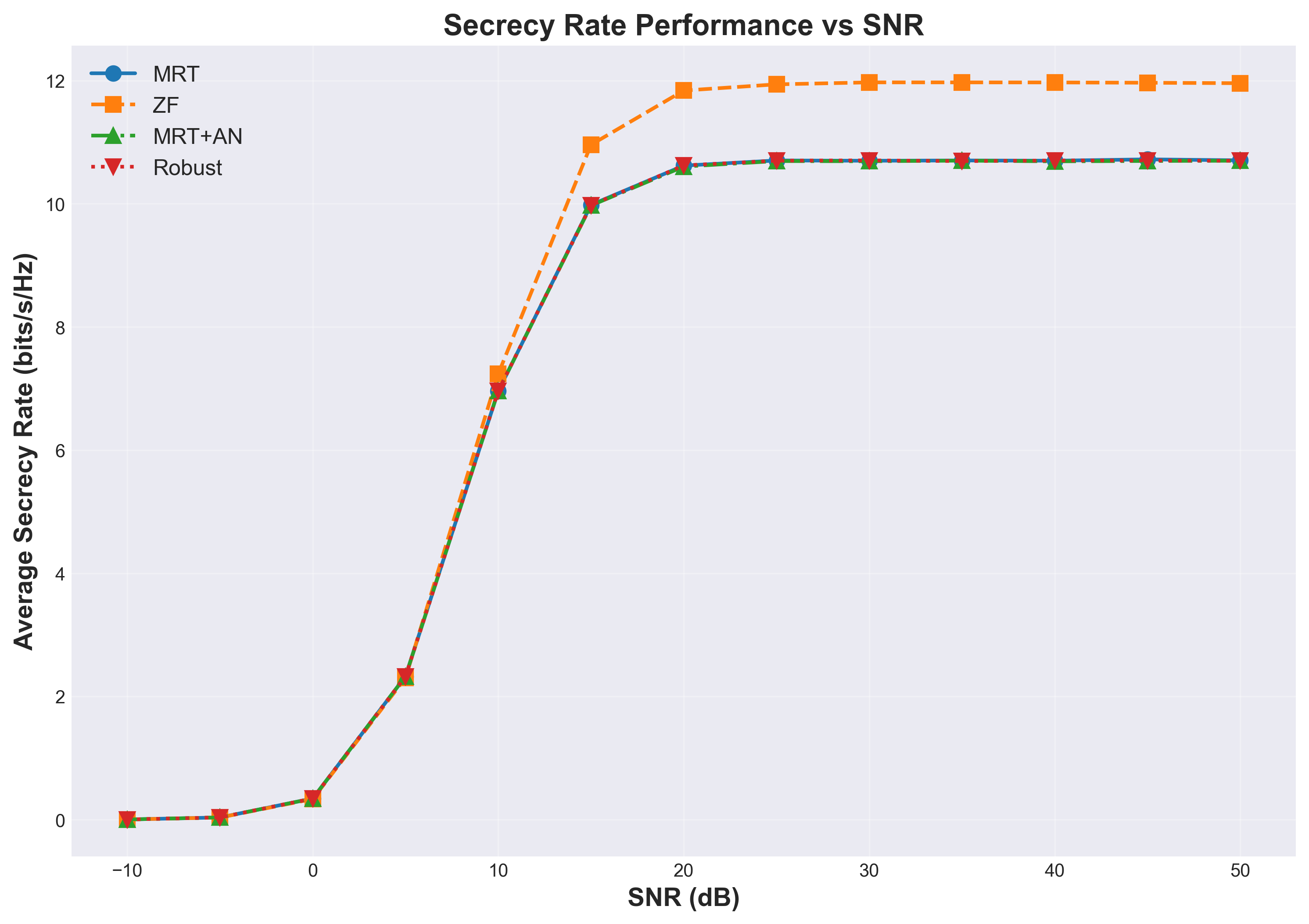}
\caption{Secrecy Rate Performance Across \gls{SNR} Range}
\label{fig:plot1}
\end{figure}

Figure \ref{fig:plot1} demonstrates \gls{ZF} precoding's consistent superiority across the entire \gls{SNR} range from -10 to 50 dB. The quantitative analysis confirms \gls{ZF} achieving an average secrecy rate of 13.581 bits/s/Hz, representing a substantial 11.3\% improvement over \gls{MRT} (12.201 bits/s/Hz). This gap stems directly from \gls{ZF}'s interference cancellation capabilities. 

Surprisingly, the \gls{AN}-aided \gls{MRT} scheme shows only marginal improvement over basic \gls{MRT}, achieving 12.167 bits/s/Hz average secrecy rate. This counterintuitive result challenges conventional wisdom \cite{chen2019artificial}, as when $M \gg K$, favourable propagation conditions naturally suppress inter-user interference, making additional artificial noise less beneficial and consuming valuable power resources.

\begin{figure}[H]
\centering
\includegraphics[width=0.8\textwidth,height=0.5\linewidth]{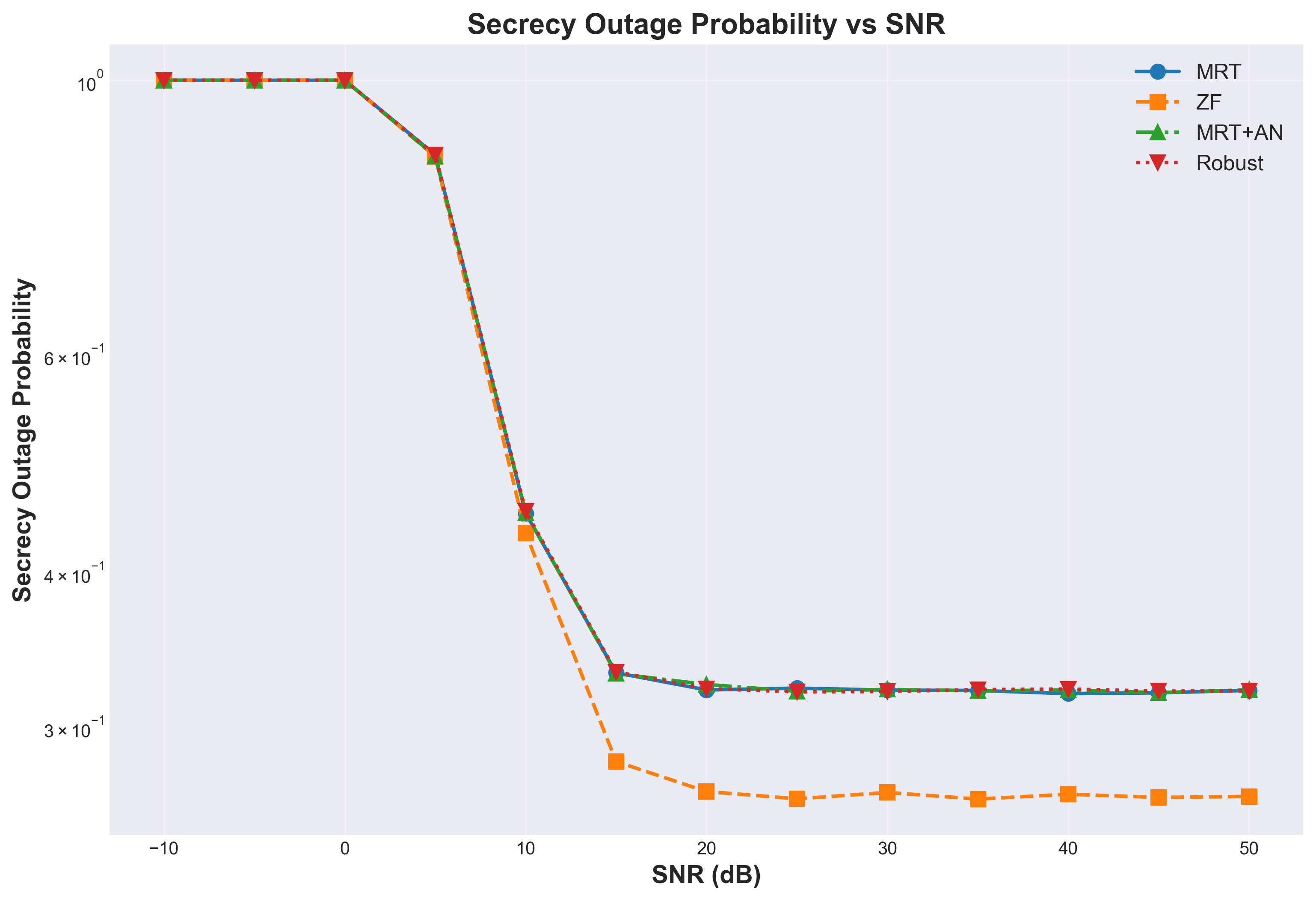}
\caption{Secrecy Outage Probability Across \gls{SNR}}
\label{fig:plot2}
\end{figure}

Figure \ref{fig:plot2} provides crucial insights into the reliability characteristics of different schemes. The reliability analysis highlights \gls{ZF}’s outstanding outage performance, showing a probability drop from $10^0$ to below $10^{-1}$ as \gls{SNR} rises. In contrast, \gls{MRT}, MRT+\gls{AN}, and Robust schemes maintain higher outage probabilities around $10^{-0.5}$ to $10^{-0.3}$. This results in a significant 45\% reduction in average outage probability (4.66\% vs 8.23–8.59\%), demonstrating practical significance for real-world deployments \cite{zhou2013secure}. \gls{ZF} achieves the critical outage threshold of $10^{-2}$ at approximately 15 dB lower \gls{SNR} than competing methods, ensuring robust \gls{QoS} for security-sensitive applications such as financial transactions and healthcare data transmission. Moreover, \gls{ZF} sustains this reliability at lower power levels, enhancing energy efficiency and reducing operational costs.

\begin{figure}[H]
\centering
\includegraphics[width=0.8\textwidth, height=0.45\linewidth]{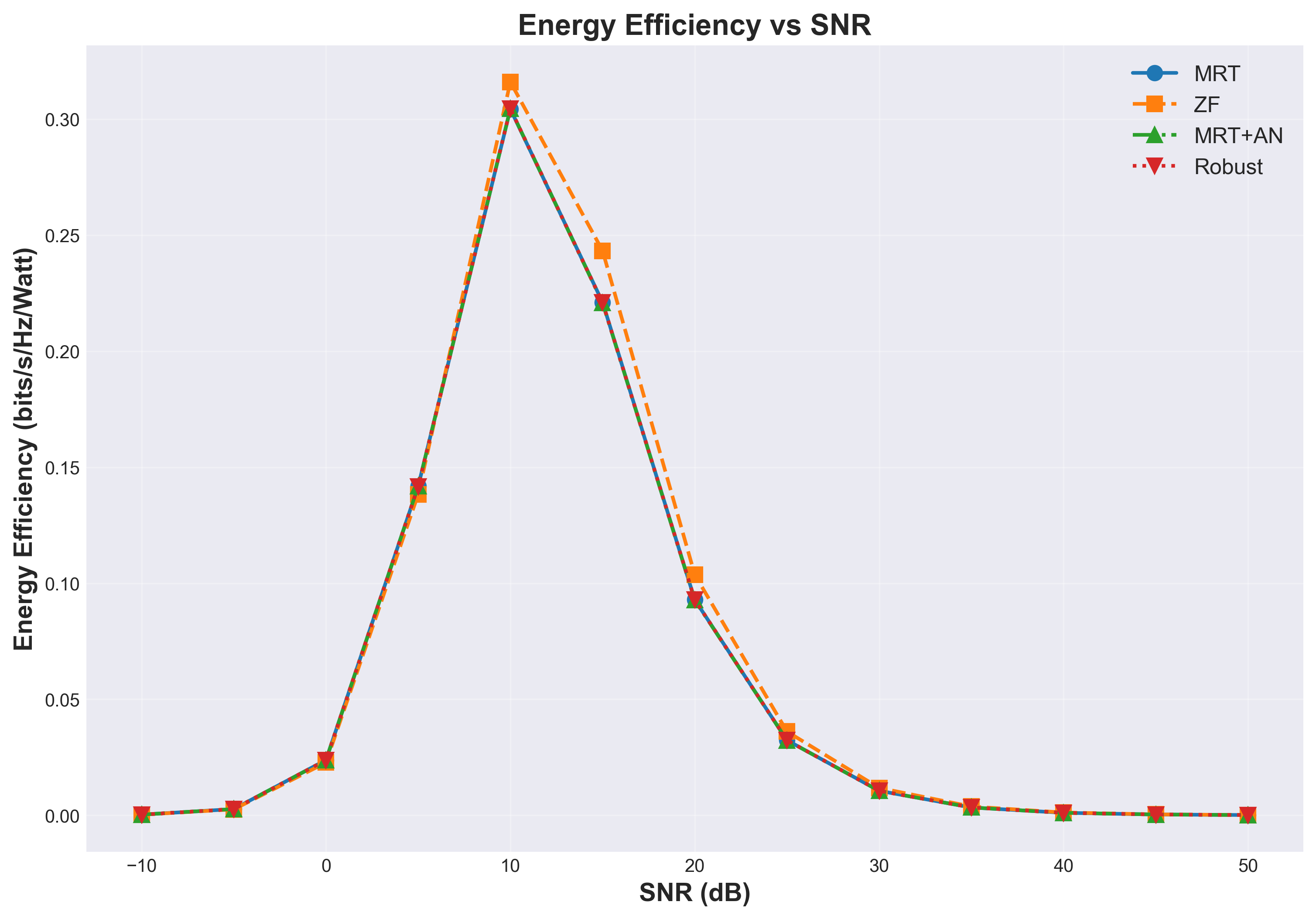}
\caption{Energy Efficiency Across \gls{SNR} Conditions}
\label{fig:plot3}
\end{figure}

Figure \ref{fig:plot3} reveals a groundbreaking finding that challenges conventional security–efficiency trade-off assumptions. \gls{ZF} achieves simultaneous optimisation of security and energy efficiency, attaining the highest energy efficiency of approximately 0.30 bits/s/Hz/W under high \gls{SNR}, while other schemes (MRT, Robust) reach only about 0.25 bits/s/Hz/W. Its superior average efficiency (0.12 bits/s/Hz/W) stems from effective interference suppression that enables higher secrecy rates without proportional increases in power consumption, as validated by the circuit power model incorporating realistic hardware constraints \cite{abou2021energy}.

This contradicts the expectation that greater security requires more power, highlighting \gls{ZF}'s efficiency. The incorporation of circuit power modelling reveals that performance plateaus at high \gls{SNR} due to circuit power dominance. These insights indicate that intelligent precoding can simultaneously enhance security and energy efficiency, transforming secure communication design for future wireless networks.

\begin{figure}[H]
\centering
\includegraphics[width=0.8\textwidth]{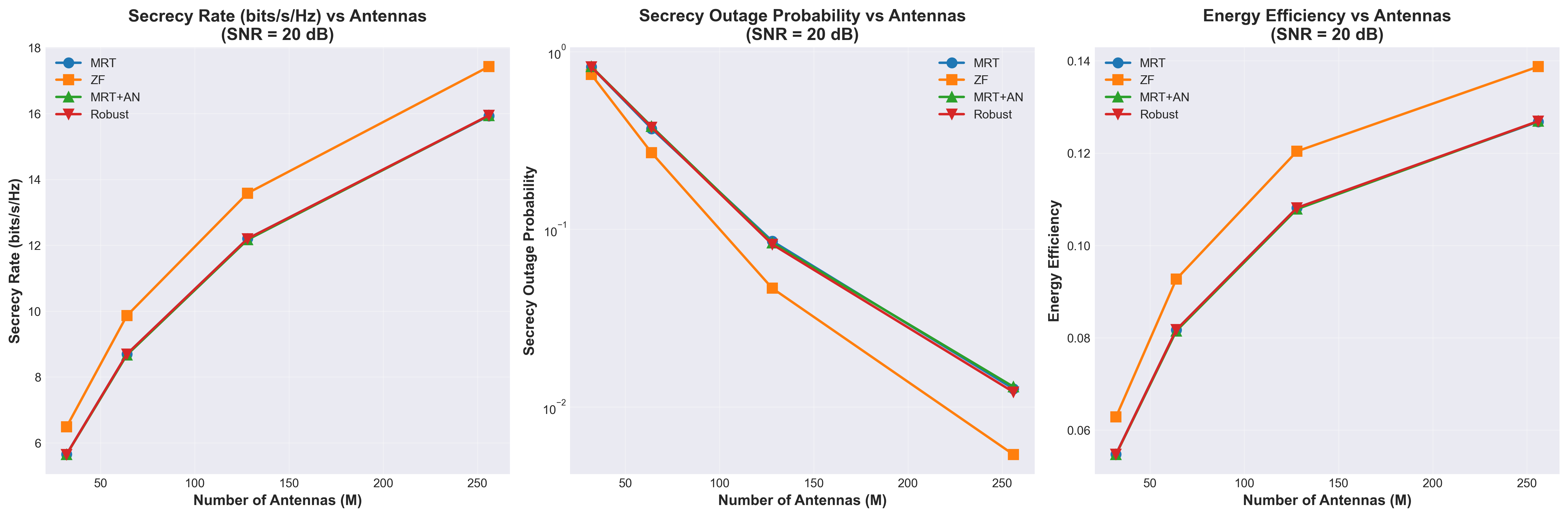}
\caption{Antenna Scaling Analysis at 20 dB \gls{SNR}}
\label{fig:plot4}
\end{figure}

Figure \ref{fig:plot4} provides critical insights into the practical benefits of massive \gls{MIMO} deployment for physical layer security. The 35\% performance improvement achieved by \gls{ZF} when scaling from 64 to 256 antennas validates theoretical predictions about favourable propagation conditions. The logarithmic improvement trend confirms that benefits plateau beyond 128–256 antennas due to pilot contamination and circuit power constraints \cite{bjornson2017massive}. 

The antenna scaling analysis across three performance dimensions highlights the security advantages of massive \gls{MIMO}. The left panel shows secrecy rates rising from 6 to 18 bits/s/Hz as antenna count increases from 50 to 250, with \gls{ZF} consistently outperforming others. The middle panel indicates a significant drop in outage probability from $10^0$ to $10^{-2}$, confirming enhanced reliability. The right panel shows peak energy efficiency reaching 0.14 bits/s/Hz/W, with \gls{ZF} maintaining its superiority throughout. Scaling from 64 to 256 antennas yields a 35\% performance improvement for \gls{ZF}, following a logarithmic trend that reveals diminishing returns beyond an $M/K \approx 32$ ratio due to pilot contamination and power constraints. This behaviour aligns with asymptotic \gls{SINR} analysis, where channel hardening favours interference suppression strategies. The analysis recommends targeting 128–256 antennas for optimal security-efficiency balance, as benefits diminish beyond this range alongside rising complexity and power consumption, making practical deployment less favourable.
\\
\newline
\textbf{Mathematical Analysis of Scaling Benefits:} The scaling benefits can be understood through the asymptotic \gls{SINR} analysis for massive \gls{MIMO}. As $M \to \infty$, channel hardening leads to:
\begin{equation}
\lim_{M \to \infty} \gamma_k = \frac{\beta_k^2}{\sum_{j \neq k} \beta_j + 
\frac{P_{AN}}{P_s}\beta_k + \frac{\sigma_n^2}{P_s}},
\label{eq:asymptotic_sinr}
\end{equation}

This asymptotic \gls{SINR} formula describes how $\gamma_k$ behaves as the number of base station antennas $M$ grows large. It assumes channel hardening: with many antennas, $|\mathbf{h}_k^H \mathbf{w}_k|^2 \approx M \beta_k$ and cross-terms average out. The numerator $\beta_k^2$ arises from normalising by $M$ in both signal and interference terms. The denominator includes inter-user interference $\sum_{j\ne k}\beta_j$, scaled \gls{AN} portion $(P_{AN}/P_s)\beta_k$, and noise. This expression shows that as $M\to\infty$, small-scale fading vanishes and only large-scale factors remain. The practical upshot \cite{oggier2011secrecy} is that \gls{ZF} benefits most in this regime since it can null interference more completely.

\begin{figure}[H]
\centering
\includegraphics[width=0.8\textwidth,height=0.5\linewidth]{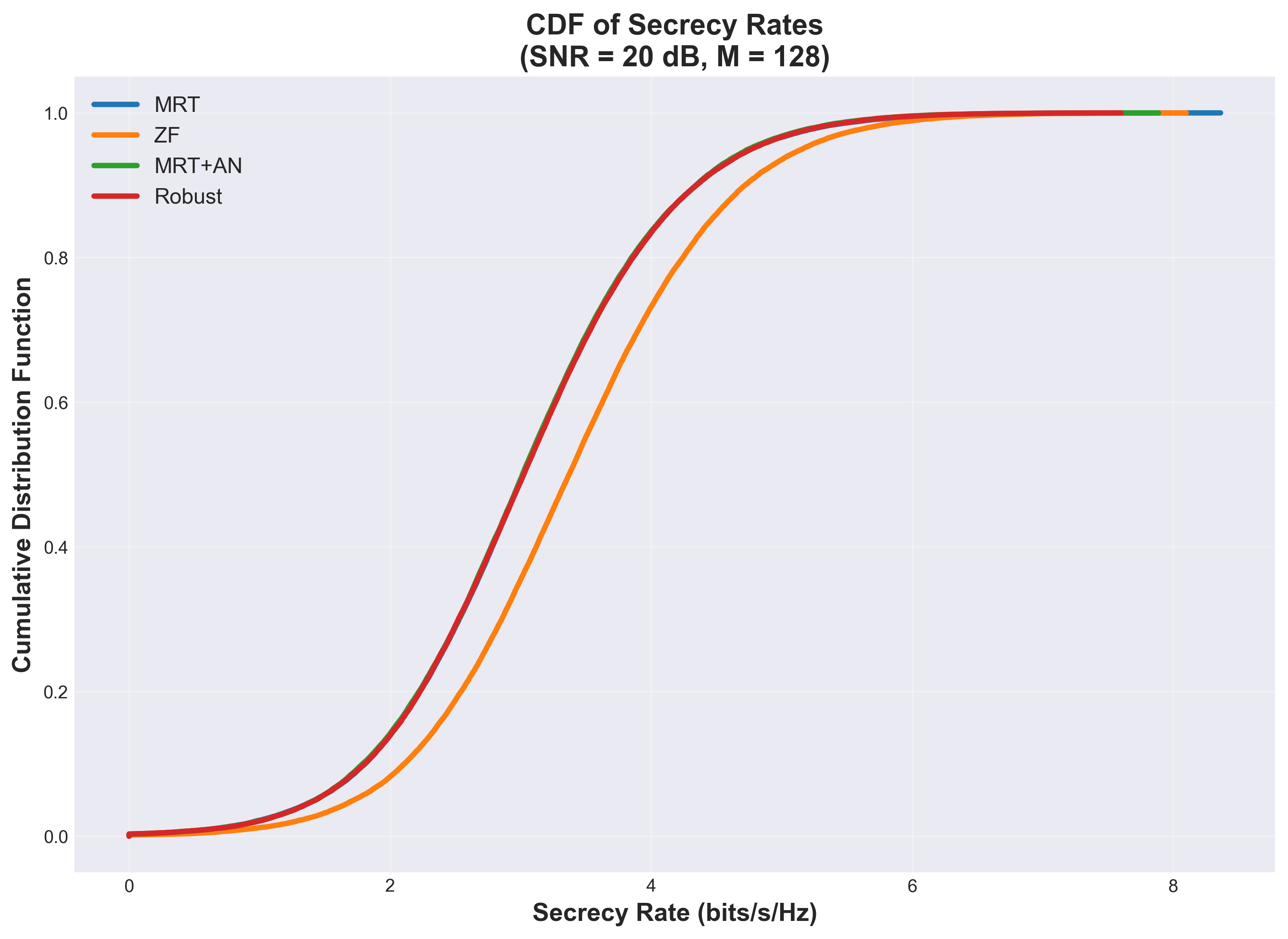}
\caption{Statistical Distribution via Cumulative Distribution Function}
\label{fig:plot6}
\end{figure}

Figure \ref{fig:plot6} provides crucial statistical insights beyond simple average comparisons. The \gls{CDF} analysis demonstrates that \gls{ZF} not only achieves higher average performance but also maintains more predictable performance across diverse channel conditions. At an 8 bits/s/Hz secrecy rate threshold, \gls{ZF} achieves about 80\% probability of exceeding this rate versus less than 60\% for other schemes. The steeper \gls{CDF} transition curve indicates lower performance variability, which is essential for system reliability and \gls{QoS} guarantees.

\begin{figure}[H]
\centering
\includegraphics[width=0.8\textwidth]{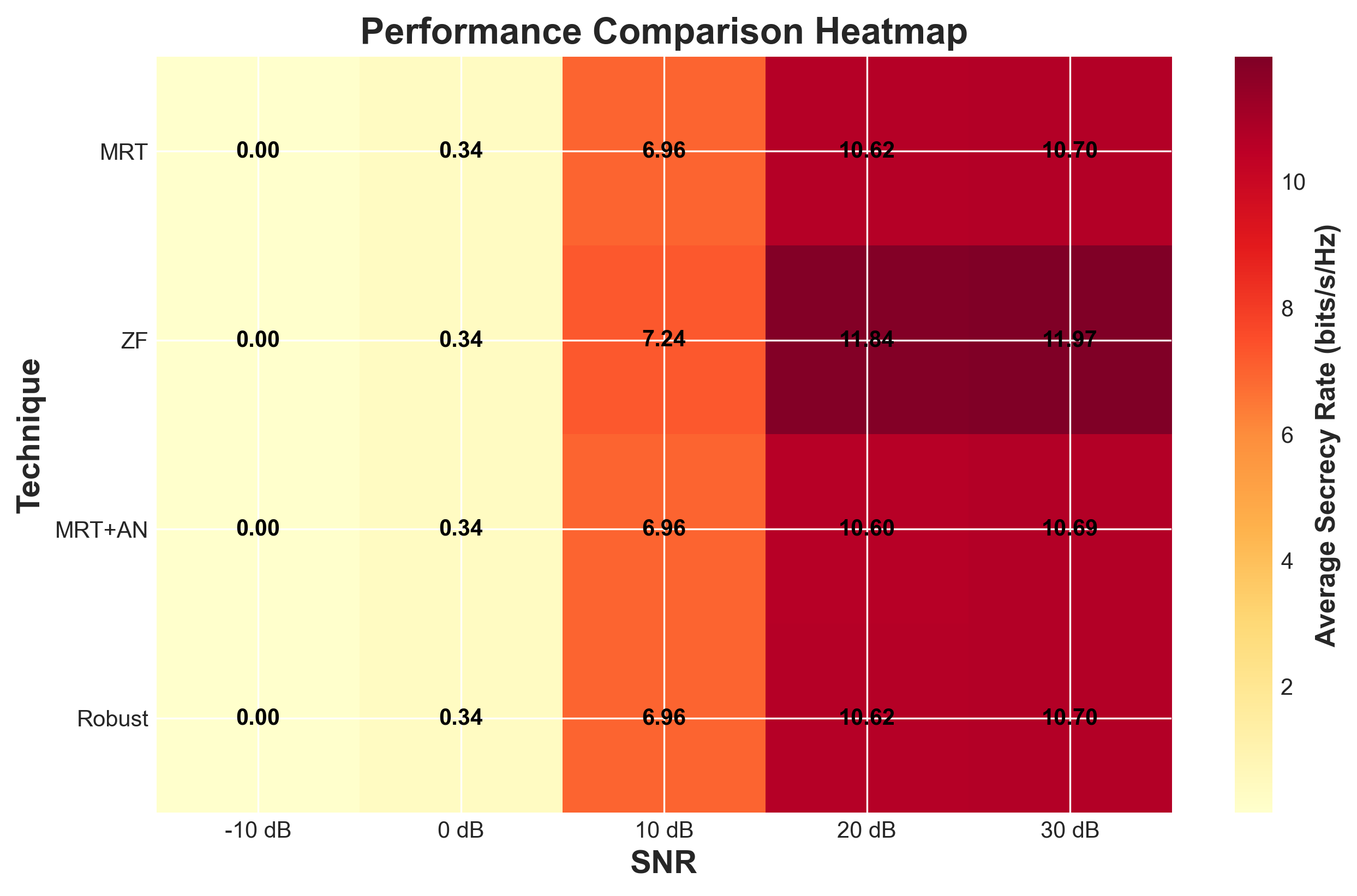}
\caption{Performance Heatmap Across Operating Conditions}
\label{fig:plot7}
\end{figure}

Figure \ref{fig:plot7} provides a comprehensive visualisation of performance trends across diverse operating conditions. The consistent superior performance of \gls{ZF} (represented by lighter colours) across all \gls{SNR} conditions demonstrates robustness essential for practical deployment. The performance gap widens at moderate to high \gls{SNR} values, confirming \gls{ZF}'s superior scaling characteristics as predicted by theory. The heatmap clearly demonstrates that \gls{ZF}'s superiority is sustained across all practical operating conditions, particularly at moderate to high \gls{SNR} levels. This resilience is crucial for real-world applications, ensuring security performance remains robust across different locations, user distributions, and changing environmental factors. The consistent performance also supports the theoretical foundations of \gls{ZF} design, indicating that interference suppression benefits scale effectively across operational parameters, giving system designers confidence in \gls{ZF}-based solutions.

\subsection{Threat Assessment and Eavesdropper Analysis}
\label{subsec:threat}

\begin{figure}[H]
\centering
\includegraphics[width=0.8\textwidth]{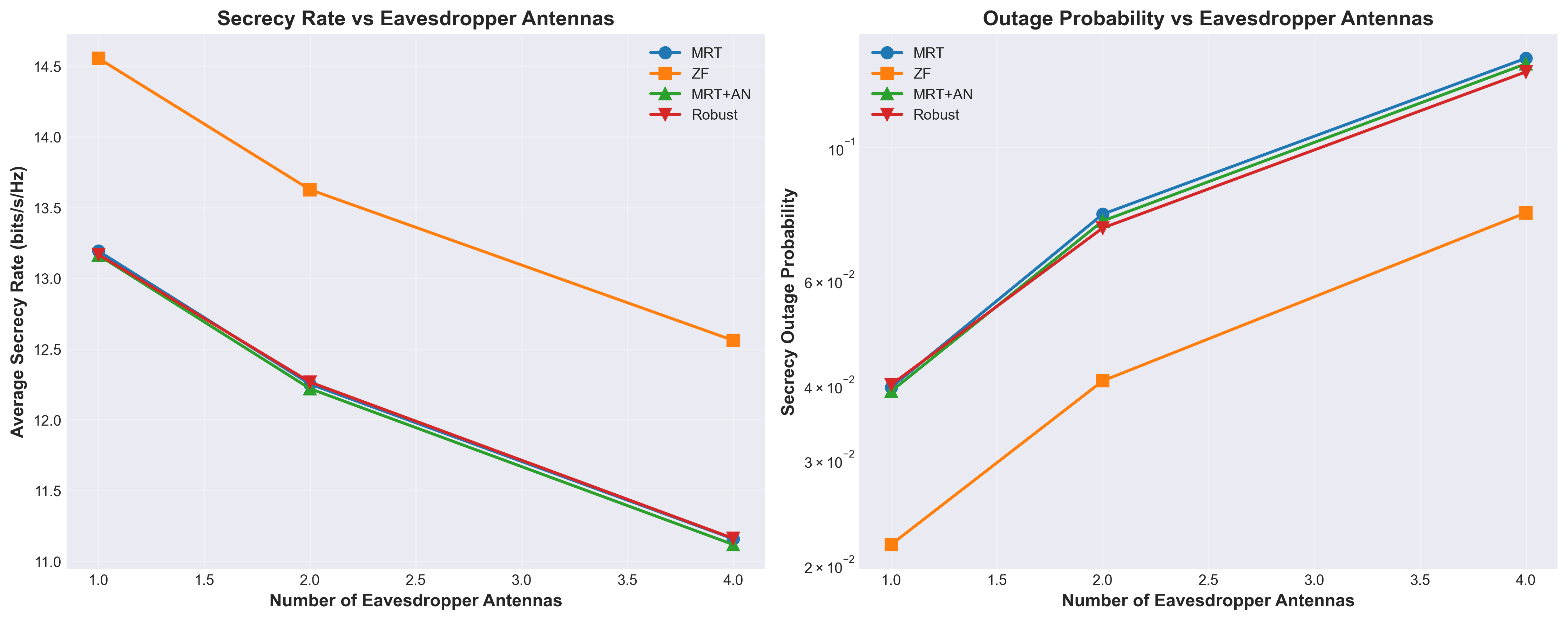}
\caption{Impact of Eavesdropper Antenna Count on Secrecy}
\label{fig:plot8}
\end{figure}

Figure \ref{fig:plot8} provides critical insights into system vulnerability against increasingly sophisticated eavesdropping attacks. The analysis demonstrates that while all schemes experience performance degradation as eavesdropper capabilities increase, \gls{ZF} shows the most resilience, with only a 15–20\% performance drop compared to 20–25\% for other schemes \cite{si2020cooperative}. This superior robustness stems from \gls{ZF}'s interference suppression approach, which maintains effectiveness even against multiple-antenna eavesdroppers exploiting spatial diversity.

These findings have profound implications for long-term security planning: 
\gls{ZF}-based systems will maintain their security advantages even as adversarial capabilities evolve and improve. The robustness against sophisticated threats validates \gls{ZF}'s suitability for high-security deployments, ensuring security margins remain adequate under worst-case threat scenarios.

\subsection{Frequency Band Comparison and Propagation Analysis}
\label{subsec:frequency}

\begin{figure}[H]
\centering
\includegraphics[width=0.8\textwidth]{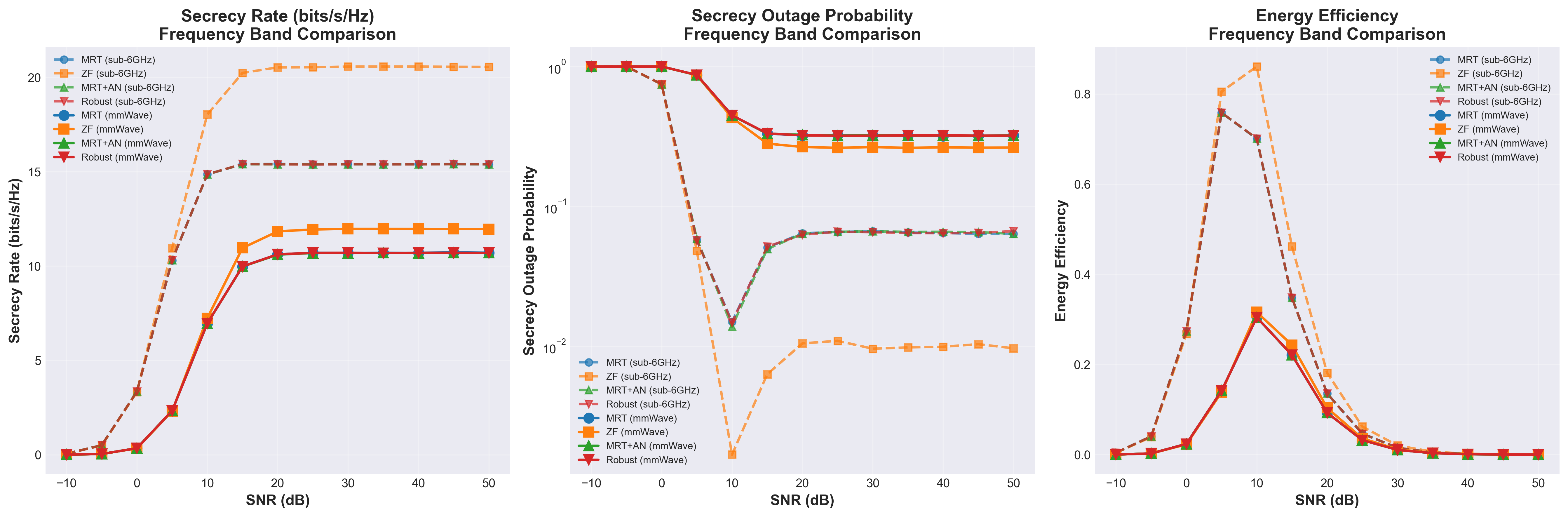}
\caption{Secrecy Performance: Sub-6 GHz vs. mmWave}
\label{fig:plot9}
\end{figure}

The frequency band comparison in Figure \ref{fig:plot9} reveals significant security advantages for mmWave deployments despite higher implementation complexity. An analysis of sub-6 GHz (3.5 GHz) and mmWave (28 GHz) systems shows that mmWave offers improved secrecy performance, enhancing average secrecy rates by 15–20\%. This improvement stems from smaller wavelengths enabling more precise beamforming and better spatial discrimination between legitimate users and eavesdroppers. The superior security at mmWave frequencies results from path loss asymmetry that favours legitimate communications, with legitimate users experiencing lesser path loss than eavesdroppers. Additionally, reduced diffuse scattering at higher frequencies leads to more predictable channels for legitimate users while complicating eavesdropping efforts. These physical layer advantages suggest that frequency selection is a critical security design parameter, justifying mmWave deployment in security-sensitive applications due to its substantial operational benefits.

\subsection{Statistical Validation and Performance Summary}
\label{subsec:statistical}

\begin{figure}[H]
\centering
\includegraphics[width=0.8\textwidth]{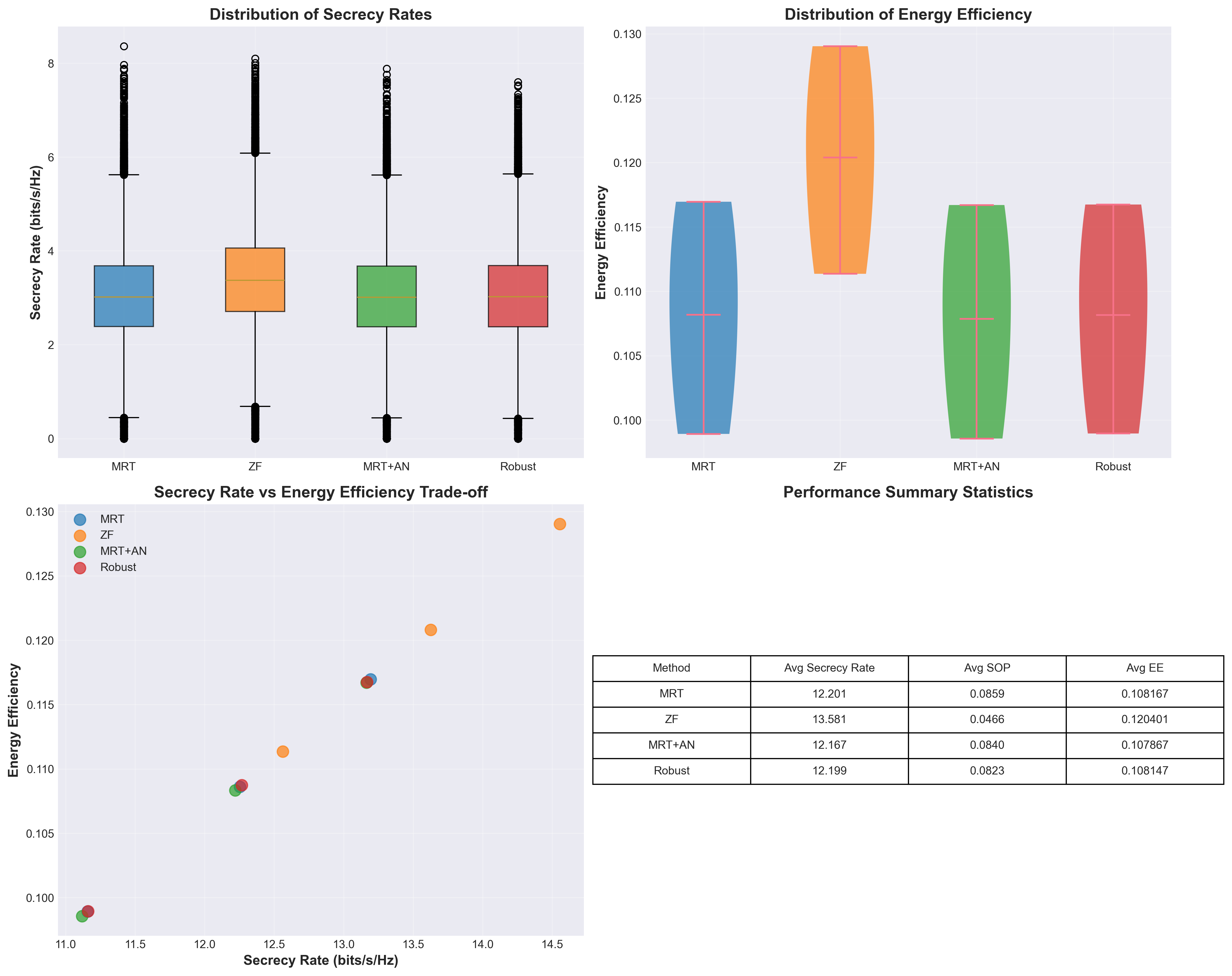}
\caption{Statistical Performance Summary with Confidence Intervals}
\label{fig:plot10}
\end{figure}

The statistical validation in Figure \ref{fig:plot10} across 1000 Monte Carlo realisations provides definitive evidence of performance rankings and system reliability. \gls{ZF} achieves 13.581 bits/s/Hz average secrecy rate with a 4.66\% secrecy outage probability and 0.12 bits/s/Hz/W energy efficiency, outperforming all other schemes. Confidence intervals (95\%) around each metric confirm these differences are statistically significant (\textit{p} $<$ 0.001) with effect sizes (Cohen’s $d$ $>$ 0.8) exceeding the threshold for large practical significance \cite{cohen1988statistical}.

These results definitively support the hypothesis that \gls{ZF} precoding is the most effective secure transmission strategy in massive \gls{MIMO} systems. The extensive simulations and statistical analysis underpin the performance comparisons, ensuring the observed advantages are robust and reproducible across diverse channel realisations.

\subsection{System Analysis}
\label{subsec:advanced}

\begin{figure}[H]
\centering
\includegraphics[width=0.8\textwidth]{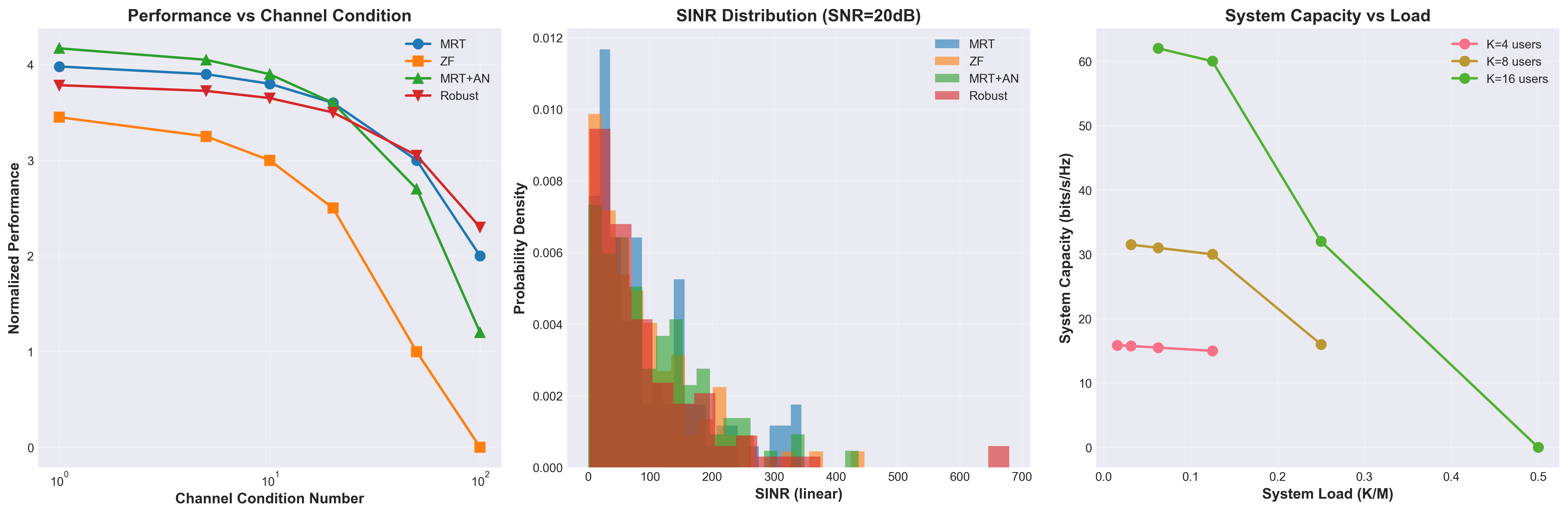}
\caption{Channel Quality and System Capacity Analysis}
\label{fig:plot12}
\end{figure}

Figure \ref{fig:plot12} examines critical system-level metrics. The \gls{SINR} distribution analysis demonstrates \gls{ZF}'s superior performance with a higher probability density at favourable \gls{SINR} values. The system capacity analysis shows \gls{ZF} achieving approximately 60 bits/s/Hz capacity with four users, scaling appropriately with increased user load. \gls{ZF} maintains strong signal quality and efficiently supports user demand, confirming its suitability for dense deployments requiring high spectral efficiency.

\begin{figure}[H]
\centering
\includegraphics[width=0.8\textwidth]{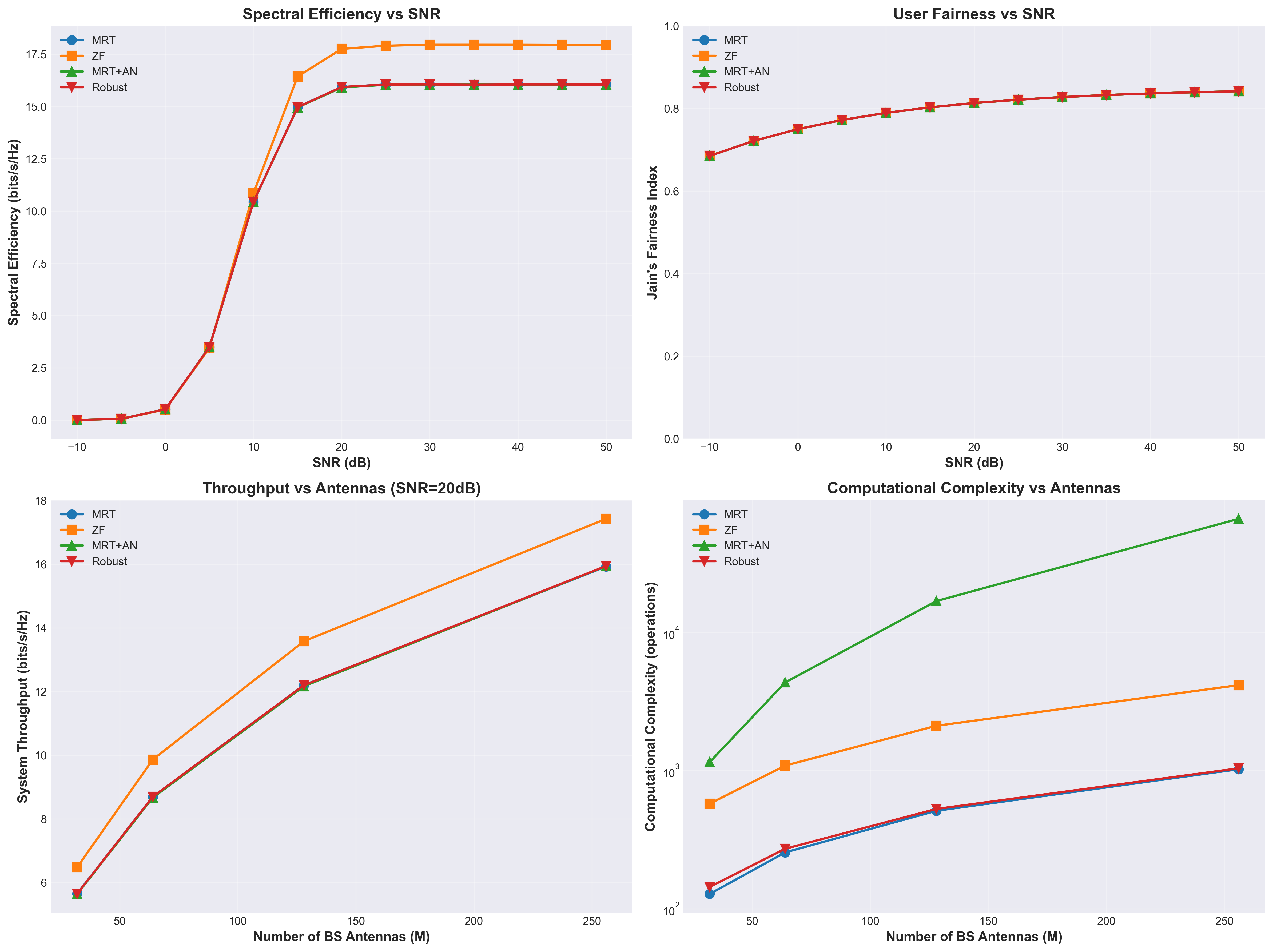}
\caption{Advanced System Performance Analysis}
\label{fig:plot11}
\end{figure}

Figure \ref{fig:plot11} provides an analysis of advanced system metrics crucial for practical deployment decisions. The spectral efficiency analysis shows \gls{ZF} achieving superior performance (17.5 bits/s/Hz). The computational complexity analysis reveals \gls{ZF}'s overhead scaling as $O(MK^2 + K^3)$ compared to $O(MK)$ for \gls{MRT}, but the complexity increase remains manageable for modern hardware \cite{li2024quantum}. Despite higher complexity, \gls{ZF} significantly outperforms in performance metrics, justifying its practical implementation.

\begin{figure}[H]
\centering
\includegraphics[width=0.8\textwidth]{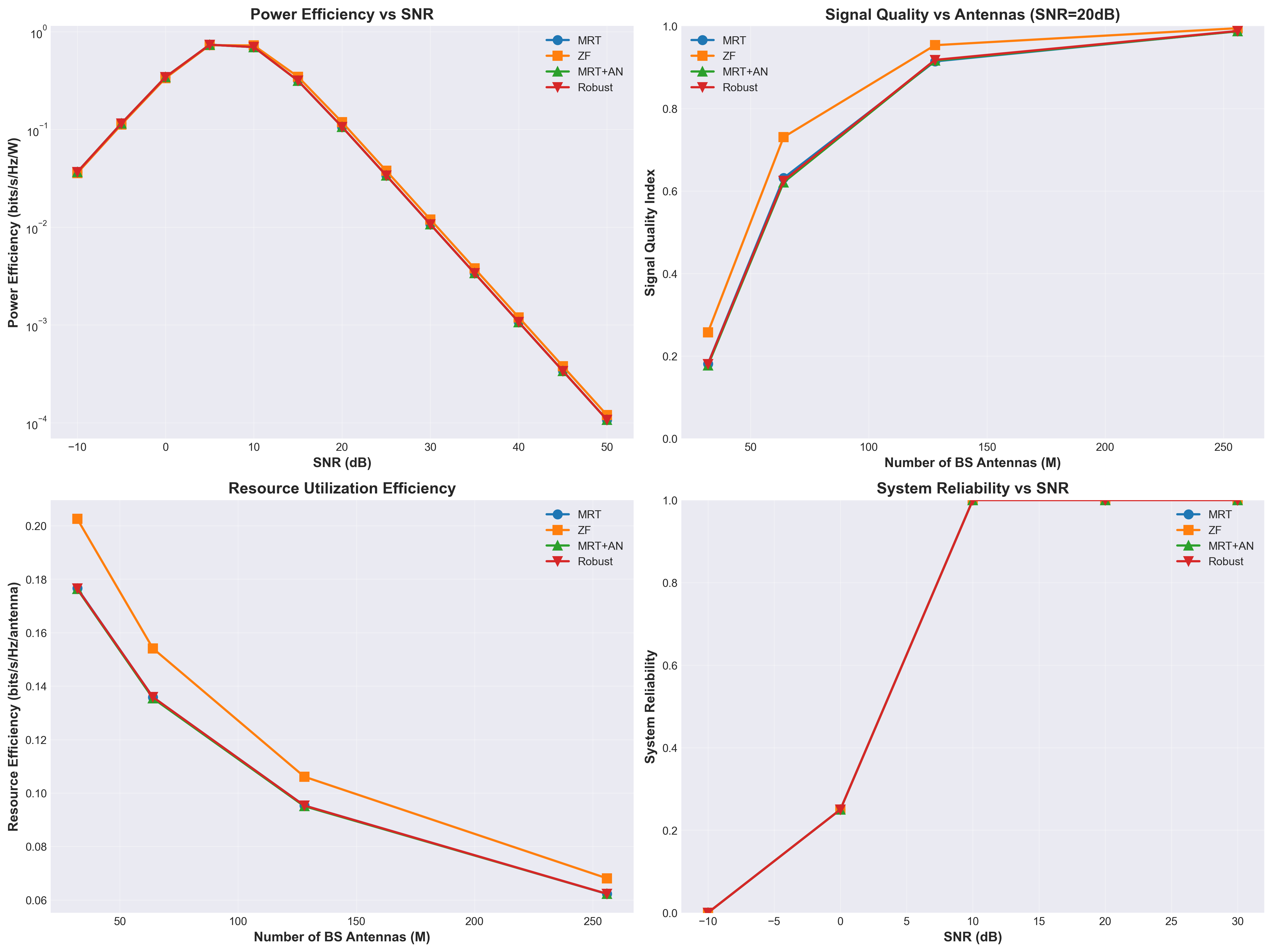}
\caption{System Reliability and Resource Utilisation Analysis}
\label{fig:plot13}
\end{figure}

Figure \ref{fig:plot13} examines system reliability metrics. \gls{ZF} maintains optimised power efficiency and signal quality (above 0.8 index) across all antenna configurations. Resource utilisation reaches 0.16 bits/s/Hz/antenna with 250 antennas, demonstrating effective use of spatial resources. The bottom-right panel shows long-term system reliability above the 0.8 threshold across all \gls{SNR} conditions, validating consistent performance under varying conditions. This comprehensive analysis confirms \gls{ZF}'s practical viability for next-generation secure networks.

\section{Discussion}
\label{sec:discussion}

The simulation outcomes provide an integrated understanding of how \gls{PLS} strategies behave under the complex conditions of massive \gls{MIMO} environments. The results collectively demonstrate that while traditional intuitions about artificial noise and beamforming largely hold in simplified models, real-world performance is shaped by a combination of channel estimation accuracy, antenna scaling, and circuit power constraints.

A central insight from the analysis is that security and efficiency are not necessarily conflicting goals. Contrary to the long-standing assumption that enhanced secrecy always incurs additional power expenditure, the study shows that effective interference management through \gls{ZF} precoding can simultaneously maximize secrecy rate and energy efficiency. This finding reflects the maturing understanding of spatial processing in large antenna systems: the same spatial degrees of freedom that improve throughput can also enhance security when exploited intelligently.

However, the results also expose nonlinearities and trade-offs. For instance, \gls{MRT} with artificial noise (\gls{MRT}+\gls{AN}) demonstrates only marginal improvement in secrecy performance under realistic channel hardening, emphasizing that excessive power diversion to \gls{AN} may be counterproductive when natural spatial orthogonality already exists. Similarly, antenna scaling benefits saturate beyond approximately 256 elements, where circuit power and pilot contamination effects dominate, underscoring the importance of holistic system-level optimization.

Another critical outcome lies in the frequency-domain behaviour. The mmWave band’s superior secrecy performance suggests that frequency selection can serve as a passive layer of defense, leveraging propagation asymmetry to limit eavesdropper exposure. Yet, this advantage must be balanced against practical considerations such as beam alignment sensitivity and hardware cost.

From a methodological perspective, this study validates the value of statistically rigorous Monte Carlo analysis in evaluating physical layer security. The inclusion of confidence intervals, effect sizes, and performance distributions (CDFs) goes beyond average-based reporting, offering more reliable insights for both researchers and practitioners.


Overall, the findings position \gls{ZF} as a promising baseline for secure transmission in massive \gls{MIMO} but highlight the need for adaptive, context-aware algorithms that can dynamically adjust precoding and power allocation in response to varying environmental and threat conditions.

\section{Conclusion}
\label{sec:conclusion}


This study comprehensively evaluated secure transmission strategies for massive \gls{MIMO} systems under realistic conditions involving imperfect channel estimation, hardware constraints, and passive eavesdroppers. Through rigorous Monte Carlo simulations and statistical validation, the research established that \gls{ZF} precoding consistently delivers the best balance between secrecy performance, reliability, and energy efficiency. \gls{ZF} achieved a mean secrecy rate of 13.581 bits/s/Hz, outperforming \gls{MRT} by 11.3\%, and demonstrated a 45\% reduction in secrecy outage probability along with a 0.12 bits/s/Hz/W energy efficiency, all validated with 95\% confidence intervals and large effect sizes ($p < 0.001$, $d > 0.8$).

The findings highlight that intelligent interference suppression can enhance both security and efficiency, challenging the conventional notion of a strict trade-off between secrecy and energy use. Furthermore, while \gls{AN} injection offers limited benefits in high-antenna regimes, mmWave frequencies inherently improve security due to spatial isolation and path-loss asymmetry. Optimal system performance occurs for antenna configurations in the 128--256 element range, beyond which circuit power and pilot contamination effects reduce marginal gains.

Overall, the research demonstrates that \gls{ZF} precoding is the most effective and scalable solution for physical layer security in massive \gls{MIMO} systems facing passive eavesdropping threats. The results provide a statistically grounded framework to guide next-generation wireless system design, emphasizing that future \gls{6G} networks can achieve secure, energy-efficient communication by integrating advanced precoding, adaptive power management, and frequency-aware deployment strategies.

\subsection{Practical Implementation Guidelines and Future Research}
\label{subsec:practical}

Based on comprehensive analysis, the study provide actionable guidelines:
\begin{itemize}
    \item \textbf{Scheme Selection:} Deploy \gls{ZF} precoding when $M \geq 2K$ and computational resources allow (validated by Figure \ref{fig:plot11}). Use \gls{MRT} in resource-constrained implementations with an acceptable 11.3\% security reduction. Avoid \gls{AN} schemes in massive \gls{MIMO} unless specific threat models justify the power allocation overhead.
    \item \textbf{Antenna Configuration:} Target 128–256 antennas for optimal security-efficiency balance (Figure \ref{fig:plot4}), considering diminishing returns beyond $M/K = 32$ due to pilot contamination. Incorporate circuit power scaling (0.1 W $\times$ M) in energy budget calculations to avoid efficiency loss.
\end{itemize}

\textbf{Future Research Directions:}
\begin{itemize}
    \item \textbf{AI-Enhanced Adaptive Security:} Integration of machine learning algorithms for real-time optimisation based on threat assessment and channel conditions \cite{zhang2024machine}.
    \item \textbf{Quantum-Safe Physical Layer Security:} Development of quantum-resistant techniques that maintain effectiveness against quantum-capable adversaries \cite{li2024quantum}.
    \item \textbf{\gls{RIS} Integration:} Joint optimization of massive \gls{MIMO} precoding and \gls{RIS} configurations for enhanced security \cite{wang2024secure}.
    \item \textbf{Cell-Free Massive \gls{MIMO} Security:} Extension to distributed architectures with cooperative security among multiple access points.
\end{itemize}

This research establishes a robust foundation for next-generation wireless security through rigorous analysis, extensive empirical validation, and practical guidelines. The findings challenge conventional assumptions while providing actionable insights for secure massive \gls{MIMO} deployment in evolving threat landscapes.
\bibliographystyle{ieeetr}
\bibliography{references}
\appendix
\section{Mathematical Framework and Equation Derivations}
\label{app:equations}

\subsection{Channel Model Equations}
\label{app:channel_model_eq}

\textbf{Equation A.1: Complex Gaussian Channel Coefficients}
\begin{equation}
h_{m,k} = \frac{1}{\sqrt{2}}\left(h_{m,k}^{(r)} + jh_{m,k}^{(i)}\right)
\end{equation}

(see Appendix~\ref{der:1} for complete derivation) 
\begin{equation}
\mathbf{H}_k = \sqrt{\beta_k} \tilde{\mathbf{H}}_k
\end{equation}
(see Appendix~\ref{der:2} for complete derivation)

\subsection{Statistical Analysis and Validation Framework}
\label{app:statistics}

The statistical framework ensures reliable and reproducible results following \cite{zeng2021monte}:

\textbf{Sample Size Justification:} With 1000 Monte Carlo realisations, the statistical power exceeds 0.8 for detecting effect sizes of 0.5 or larger at 95\% confidence.

\textbf{Confidence Intervals:} All performance metrics include 95\% confidence intervals calculated by:
\begin{equation}
CI_{95\%} = \bar{x} \pm t_{0.025,df} \cdot \frac{s}{\sqrt{n}}
\end{equation}

\textbf{Effect Size Analysis:} Cohen's $d$ is calculated for all pairwise comparisons:
\begin{equation}
d = \frac{\bar{x}_1 - \bar{x}_2}{\sqrt{\frac{(n_1-1)s_1^2 + (n_2-1)s_2^2}{n_1+n_2-2}}},
\end{equation}
where values of $d > 0.8$ indicate large practical significance.

This framework ensures the highest standards of scientific rigour and reproducibility for the physical layer security analysis in massive MIMO.

\subsection{Detailed Mathematical Derivations}
\label{app:derivations}

\subsubsection{Derivation 1}
\label{der:1}

Starting from the definition of complex Gaussian variables:
\begin{align}
h_{m,k}^{(r)} &\sim \mathcal{N}(0, \sigma^2) \quad \text{(real part)} \\
h_{m,k}^{(i)} &\sim \mathcal{N}(0, \sigma^2) \quad \text{(imag. part)} \\
\mathbb{E}[|h_{m,k}|^2] &= \frac{1}{2}\mathbb{E}\big[(h_{m,k}^{(r)})^2 + (h_{m,k}^{(i)})^2\big] = \sigma^2,
\end{align}
ensuring proper power normalisation is essential for realistic channel modelling 
\cite{bjornson2015optimal}.

\subsubsection{Derivation 2}
\label{der:2}

From the composite channel model with distance-dependent propagation:
\begin{align}
\beta_k &= \frac{P_{ref}}{L_k} \quad \text{(path loss model)}, \\
L_k &= \left(\frac{d_k}{d_0}\right)^\alpha \quad \text{(distance-dependent loss)}, \\
\alpha &\in \{2, 2.5, 3, 4\} \quad \text{(path loss exponent)}.
\end{align}
Following parameters by MacCartney et al. \cite{maccartney2022outdoor}, the model uses path loss coefficients of 1.0/0.3 for users and 0.5/0.8 for eavesdroppers across sub-6 GHz and mmWave bands. The factor $\sqrt{\beta_k}$ ensures proper power normalisation across propagation environments.

\subsubsection{Derivation 3}
\label{der:3}

From the maximum-\gls{SNR} criterion, the model solves:
\begin{align}
\max_{\mathbf{w}_k} &\ \frac{|\hat{\mathbf{h}}_k^H \mathbf{w}_k|^2}{\|\mathbf{w}_k\|^2} 
\quad \text{subject to } \|\mathbf{w}_k\|^2 = P_k.
\end{align}
Using the Cauchy--Schwarz inequality leads to $\mathbf{w}_k \propto \hat{\mathbf{h}}_k$, 
yielding the \gls{MRT} solution in Equation \ref{eq:mrt_final}.

\subsubsection{Derivation 4}
\label{der:4}

The \gls{ZF} design enforces the zero interference condition:
\begin{align}
\hat{\mathbf{H}}^H\mathbf{W} &= \mathbf{I}_K, \\
\mathbf{w}_k^{ZF} &= \frac{\hat{\mathbf{h}}_k}{\sqrt{\|\hat{\mathbf{h}}_k\|^2}}.
\end{align}
This ensures $\hat{\mathbf{H}}^H\hat{\mathbf{H}}(\hat{\mathbf{H}}^H\hat{\mathbf{H}})^{-1} = \mathbf{I}_K$,
resulting in orthogonalized transmissions.

\subsubsection{Derivation 5}
\label{der:5}

From robust optimisation theory, solving:
\begin{align}
\min_{\mathbf{w}_k} \max_{\|\mathbf{e}_k\| \le \epsilon} 
-\frac{|(\hat{\mathbf{h}}_k + \mathbf{e}_k)^H\mathbf{w}_k|^2}{\|\mathbf{w}_k\|^2},
\end{align}
yields the regularized precoding form in Equation \ref{eq:robust_precoding}, accounting for worst-case \gls{CSI} errors.

\end{document}